\documentclass[a4paper,12pt]{article}
\usepackage{graphics}
\usepackage{amsfonts}

\begin{document}
%

\newcommand{\be}{\begin{equation}}
\newcommand{\ee}{\end{equation}}
\newcommand{\bea}{\begin{eqnarray}}
\newcommand{\eea}{\end{eqnarray}}
\newcommand{\bean}{\begin{eqnarray*}}
\newcommand{\eean}{\end{eqnarray*}}
\font\upright=cmu10 scaled\magstep1
\font\sans=cmss12
\newcommand{\ssf}{\sans}
\newcommand{\stroke}{\vrule height8pt width0.4pt depth-0.1pt}
\newcommand{\Z}{\hbox{\upright\rlap{\ssf Z}\kern 2.7pt {\ssf Z}}}
\newcommand{\ZZ}{\Z\hskip -10pt \Z_2}
\newcommand{\C}{{\rlap{\upright\rlap{C}\kern 3.8pt\stroke}\phantom{C}}}
\newcommand{\R}{\hbox{\upright\rlap{I}\kern 1.7pt R}}
\newcommand{\HH}{\hbox{\upright\rlap{I}\kern 1.7pt H}}
\newcommand{\CP}{\hbox{\C{\upright\rlap{I}\kern 1.5pt P}}}
\newcommand{\identity}{{\upright\rlap{1}\kern 2.0pt 1}}
\newcommand{\half}{\frac{1}{2}}
\newcommand{\pr}{\partial}
\newcommand{\bm}{\boldmath}
\newcommand{\I}{{\cal I}} 
\newcommand{\M}{{\cal M}}
\newcommand{\N}{{\cal N}}
\newcommand{\e}{\varepsilon}

\thispagestyle{empty}
\rightline{DAMTP-2010-9}
\vskip 5em
\begin{center}
{{\bf \Large Maximally Non-Abelian Vortices from \\
Self-dual Yang--Mills Fields
}} 
\\[18mm]

{\normalsize
Nicholas S.~Manton$^{a}$ 
and
 Norisuke~Sakai$^b$}
\footnotetext{
email \tt
N.S.Manton(at)damtp.cam.ac.uk;
sakai(at)lab.twcu.ac.jp.
}

\vskip 1.5em
$^a$ {\it 
Department of Applied Mathematics and Theoretical Physics,\\
University of Cambridge, Wilberforce Road, Cambridge CB3 0WA, U.K.}
\\
$^b$ {\it 
Department of Mathematics, Tokyo Woman's Christian University, 
Zenpukuji, Suginami, 
Tokyo 167-8585, Japan }
 \vspace{12mm}

\abstract{
A particular dimensional reduction of $SU(2N)$ Yang--Mills theory on
$\Sigma \times S^2$, with $\Sigma$ a Riemann surface, yields an
$S(U(N) \times U(N))$ gauge theory on $\Sigma$, with a matrix Higgs field.
The $SU(2N)$ self-dual Yang--Mills equations reduce to
Bogomolny equations for vortices on $\Sigma$. These equations are
formally integrable if $\Sigma$ is the hyperbolic plane, and we
present a subclass of solutions.
}

\end{center}

\vfill
\newpage
\setcounter{page}{1}
\setcounter{footnote}{0}
\renewcommand{\thefootnote}{\arabic{footnote}}


\noindent
{\bf 1.\ Introduction} 

The generalization of abelian Higgs vortices to the 
non-abelian case has recently gained much attention 
\cite{Hanany:2003hp,Auzzi:2003fs,Eto:2005yh}. There are many 
variants of non-abelian vortices, and in this
paper we shall investigate one of these, one that has not been
explicitly investigated before, but which has a mathematically elegant
and symmetric structure. All these types of vortices satisfy
static, first order Bogomolny equations, defined in two-dimensional
space. Vortices are most commonly studied on the
plane $\R^2$, but the Bogomolny equations are not integrable
there. The vortex equations on the hyperbolic plane $\HH^2$
are, however, integrable \cite{Witten:1976ck, Leznov:1980tz, Popov}. 
The reason is that these vortex
equations arise by dimensional reduction of the self-dual Yang--Mills
equations on $\HH^2 \times S^2$, where the curvatures on $\HH^2$ and
the 2-sphere $S^2$ are opposite; moreover there is a conformal equivalence 
$\HH^2 \times S^2 \cong \R^4 - \R^1$, and self-dual Yang--Mills is both 
conformally invariant, and integrable on $\R^4$. 
The vortex equations on $\R^2$ also arise by
dimensional reduction of self-dual Yang--Mills, this time on 
$\R^2 \times S^2$, but here there is no integrability. Solutions exist
despite this, but they are transcendental, and their existence has to
be established by methods of analysis, or numerics \cite{Taubes}.

The dimensional reduction leading from self-dual Yang--Mills fields 
to vortices arises by imposing spherical symmetry (i.e. $SO(3)$
symmetry) on the gauge field over the $S^2$ factor of a Riemannian product
4-manifold $\Sigma \times S^2$, where $\Sigma$ is a Riemann surface. 
The resulting vortex equations are on $\Sigma$. Since $SO(3)$ is 
non-abelian, the dimensional reduction is non-trivial,
and there are various possible outcomes. Spherically symmetric $SU(2)$
gauge fields were first presented in the 1970's in the context of
monopoles and instantons. A systematic understanding was achieved by
Romanov et al. \cite{RomST,Man}, and a more general overview of 
symmetric gauge fields was given in ref. \cite{ForMan}. The
mathematical basis for this can be traced back to the earlier theorem 
of Wang \cite{Wang}, but the later work incorporated
dynamical aspects like the Yang--Mills action and field equations.

We will briefly review the general structure of $SO(3)$-symmetric pure 
Yang--Mills fields with gauge group ${\cal G}$ on $\Sigma \times
S^2$, and show that the dimensionally reduced self-dual Yang--Mills
equations are Bogomolny equations for vortices on $\Sigma$, with a 
gauge group $G$ that is a subgroup of ${\cal G}$. 
We then focus on an example where $G$ is a particularly 
large subgroup of ${\cal G}$. Here ${\cal G} = SU(2N)$ and 
$G = S(U(N) \times U(N))$. This is at the opposite extreme 
from another well-studied case, where $G$ is particularly
small, namely $G = U(1)^{2N-1}$ \cite{LezSav,Bais:1978ex,Wilkinson:1978zh}.

The Bogomolny equations on $\Sigma$ involve a
$G$-gauge potential and also Higgs fields. The latter arise from the
components of the original ${\cal G}$-gauge potential tangent (more
accurately, co-tangent) to $S^2$. In our example, the Higgs field is a
complex $N \times N$ matrix, gauge transforming from the left and
right by the two $U(N)$ factors of $G$. Our example is
therefore closely related to the well known non-abelian vortex
equations with an $N_c \times N_f$ matrix of Higgs fields, where there
is a ``colour'' $U(N_c)$ gauge group acting from the left, and a ``flavour''
$SU(N_f)$ global symmetry group acting from the right.  These
colour-flavour theories arise naturally in supersymmetric gauge
theories with eight supercharges
\cite{ShiYung}. It is usually assumed that $N_f \ge N_c$, to 
have a vacuum solution of zero energy, where the colour and 
the flavour are locked together. 

We will present our Bogomolny equations for both $\Sigma = \R^2$ and 
$\Sigma = \HH^2$. One Bogomolny equation implies that in a
certain sense the Higgs field is holomorphic. The free parameters of
the holomorphic Higgs field are the moduli of the vortex solutions. 
The other Bogomolny equations then reduce to gauge-invariant 
``master equations'', a generalization of Taubes' equation for abelian 
vortices \cite{Taubes}. It is expected that the master equations have unique
solutions once the holomorphic Higgs field is fixed. In the
hyperbolic case, $\Sigma = \HH^2$, the master equations simplify, and 
are formally completely integrable. However, we have not found a
general explicit solution satisfying the boundary conditions. We do
show, however, that the explicitly known hyperbolic abelian vortices,
found by Witten \cite{Witten:1976ck}, can be embedded as solutions 
in the non-abelian system. These embedded abelian vortices are 
intrinsically non-abelian, in the same sense as the 
well-known non-abelian vortices in the Higgs phase 
 \cite{Hanany:2003hp,Auzzi:2003fs,Eto:2005yh}. 

More general explicit solutions could emerge from an application of
the formulae of Leznov and Saveliev \cite{Leznov:1980tz}. These rely on a good
understanding of the structure of the gauge groups, but appear not to
incorporate boundary conditions. The twistor approach of Popov could
be useful, but so far has not yielded explicit solutions \cite{Popov}. More
promising, possibly, is the recent work of Manton and Rink, in which
hyperbolic abelian vortices are constructed in a purely geometrical
way, reproducing Witten's solutions and also giving novel
solutions on surfaces $\Sigma$, other than $\HH^2$, that have a 
hyperbolic metric \cite{ManRink}. Finding a non-abelian generalization 
of this approach would be useful and interesting. 

\vspace{5mm}

\noindent
{\bf 2.\ Self-duality and Bogomolny equations}

Bogomolny equations for vortices on a Riemann surface $\Sigma$ arise
naturally by dimensional reduction of the self-dual Yang--Mills
equations on $\Sigma \times S^2$. Let
$z$ be a complex coordinate 
on $\Sigma$, and $y$ the standard complex 
coordinate on $S^2$ obtained by stereographic projection (so that $y =
\tan\frac{\theta}{2} \, e^{i\varphi}$ with $\theta, \varphi$ usual polar
coordinates). The metric on $\Sigma \times S^2$ is taken to be 
\be
ds^2 = \sigma(z,{\bar z})dz d{\bar z} 
+ \frac{8}{(1 + y{\bar y})^2} dy d{\bar y} \,.
\label{metric}
\ee
$\sigma$ is a generic conformal factor on $\Sigma$, and the second term
describes a 2-sphere of fixed radius $\sqrt{2}$ and Gauss curvature $\half$.

Let the gauge group be ${\cal G}$, a compact Lie group with Lie algebra
${\mathfrak g}$, whose complexification is ${\mathfrak g}^*$. The 
Yang--Mills gauge potential has components ${\cal A}_z, 
{\cal A}_{\bar z}, {\cal A}_y, {\cal A}_{\bar y}$ with values in 
${\mathfrak g}^*$, but ${\cal A}_z + {\cal A}_{\bar z}$ and $i({\cal A}_z -
{\cal A}_{\bar z})$, being components in real directions, 
must be in ${\mathfrak g}$ itself \footnote{
More explicitly, if ${\cal G}$ is a group of unitary matrices, with a 
Lie algebra ${\mathfrak g}$ of antihermitian matrices, then 
${\cal A}_z + {\cal A}_{\bar z} = -({\cal A}_z + {\cal A}_{\bar
  z})^\dagger$ and ${\cal A}_z - {\cal A}_{\bar z} = ({\cal A}_z -
{\cal A}_{\bar z})^\dagger$. So ${\cal A}_z$ and ${\cal A}_{\bar z}$
are not in general antihermitian, but by adding or subtracting these
equations we see that ${\cal A}_{\bar z} = -{\cal A}_{z}^\dagger$.},
and similarly for ${\cal A}_y, {\cal A}_{\bar y}$. 

We now suppose that the gauge potential is $SO(3)$-invariant over the
2-sphere, $S^2$. $SO(3)$ does not act freely on $S^2$. The isotropy 
group at each point of $S^2$ (the subgroup keeping that
point fixed) is $SO(2)$. Let us focus on the particular point $y=0$,
and its $SO(2)$ isotropy group. For the gauge potential to be
``invariant'' at $y=0$ and its infinitesimal
neighbourhood, we mean that it is invariant under a combined $SO(2)$ rotation 
and gauge transformation. To define the gauge transformation, we must 
identify a subgroup $SO(2)_{\cal G}$ in ${\cal G}$ (which can be chosen to be 
constant over $\Sigma$). Let the
generator of $SO(2)_{\cal G}$ be denoted by $\Lambda$, such that in the
adjoint representation of ${\cal G}$, $\exp(2\pi\Lambda)$ is the
identity. The combined
action of $SO(2)$ then consists of rotations by $\alpha$ combined with gauge
transformations by $\exp(\alpha\Lambda)$, and the gauge potential must
be invariant under this. Having chosen this lift of the 
$SO(2)$-action at $y=0$, one can show that the notion of an
$SO(3)$-invariant gauge potential over $\Sigma \times S^2$ is completely fixed,
and in a convenient choice of gauge, the general invariant gauge 
potential on $\Sigma \times S^2$ is given by the formulae 
\cite{Man,Gar,Popov}
\bea
{\cal A}_z &=& {\bf A}_z(z,{\bar z}) \label{SO3inv1} \\ 
{\cal A}_{\bar z} &=& {\bf A}_{\bar z}(z,{\bar z}) \label{SO3inv2} \\
{\cal A}_y &=& \frac{1}{1 + y{\bar y}}
(-\Phi(z,{\bar z}) - i\Lambda{\bar y}) \label{SO3inv3} \\
{\cal A}_{\bar y} &=& \frac{1}{1 + y{\bar y}}
(\bar\Phi(z,{\bar z}) + i\Lambda y) \,.
\label{SO3inv4}
\eea
Here, the dependence on $z$ and ${\bar z}$ is arbitrary, but the
dependence on $y$ and ${\bar y}$ is as shown. In addition, there are
linear constraints, arising from the $SO(2)$ invariance
at $y=0$, namely
\be
[\Lambda, {\bf A}_z] = [\Lambda, {\bf A}_{\bar z}] = 0
\label{constraints1}
\ee
\be
[\Lambda, \Phi ] = -i \Phi \,, \quad [\Lambda, {\bar\Phi}] =
i{\bar\Phi} \,.
\label{constraints2}
\ee
The interpretation of these constraints is that ${\bf A}_z, 
{\bf A}_{\bar z}$ are 
components of a gauge potential on $\Sigma$ for the gauge group $G$ 
which is the centralizer of $SO(2)_{\cal G}$ in ${\cal G}$. Also, 
$\Phi,{\bar\Phi}$ are scalar Higgs fields on $\Sigma$ which must lie
in the $\mp i$ eigenspaces of ${\rm ad} \, \Lambda$ in ${\mathfrak g}^*$. These
eigenspaces are representation spaces for $G$, so $\Phi,{\bar\Phi}$
are Higgs fields transforming under these representations of $G$.

The self-dual Yang--Mills equations on $\Sigma \times S^2$, with
metric (\ref{metric}) and gauge group ${\cal G}$, are
\bea
\frac{8}{(1 + y{\bar y})^2}{\cal F}_{z{\bar z}} &=&  
\sigma{\cal F}_{y{\bar y}} \\
{\cal F}_{z{\bar y}} &=& 0 \\
{\cal F}_{{\bar z}y} &=& 0 \,,
\eea
where ${\cal F}_{\mu\nu} = \pr_{\mu} {\cal A}_{\nu} -
\pr_{\nu} {\cal A}_{\mu} + [{\cal A}_{\mu}, {\cal A}_{\nu}]$ 
for any coordinate indices $\mu, \nu$. Substituting the 
$SO(3)$-invariant fields (\ref{SO3inv1})--(\ref{SO3inv4}) into this 
set of equations yields
\bea
{\bf F}_{z{\bar z}} &=& \frac{\sigma}{8} 
\left(2i\Lambda - [\Phi, {\bar \Phi}]\right) \label{B1} \\
D_z {\bar\Phi} &=& 0 \label{B2} \\
D_{\bar z} \Phi &=& 0 \,, \label{B3} 
\eea
where ${\bf F}_{z{\bar z}} = \pr_z {\bf A}_{\bar z} - \pr_{\bar z} {\bf A}_z 
+ [{\bf A}_z, {\bf A}_{\bar z}]$, $D_z {\bar\Phi} 
= \pr_z {\bar\Phi} + [{\bf A}_z, {\bar\Phi}]$ 
and $D_{\bar z} \Phi = \pr_{\bar z} \Phi + [{\bf A}_{\bar z}, \Phi]$. 
It is consistent to interpret these as unconstrained Bogomolny equations with
gauge group $G$, and this is seen explicitly if the linear constraints 
(\ref{constraints1}) and (\ref{constraints2}) are solved. For example, 
both left and right hand sides of (\ref{B1}) are in the zero
eigenspace of ${\rm ad} \, \Lambda$, which is the Lie algebra of $G$.

We have so far presented the most general type of $SO(3)$-invariant
gauge field. There are two related reasons to restrict the choice of
$\Lambda$. The first comes from requiring that the vortex
solutions of the Bogomolny equations have finite energy. If $\Sigma$ 
has infinite area, as $\R^2$ and $\HH^2$ do, then approaching infinity 
(the boundary of $\Sigma$), the solution must approach the
vacuum. This means that ${\bf F}_{z{\bar z}} = 0$ there, and hence
\be
2i\Lambda - [\Phi, {\bar \Phi}] = 0 \,.
\label{vac}
\ee
If we denote the vacuum values of $\Phi, {\bar \Phi}$ by 
$\Phi_0, {\bar \Phi}_0$ respectively, then, combining (\ref{vac})
and the constraints (\ref{constraints2}), we have 
\be
[\Lambda, \Phi_0 ] = -i \Phi_0 \,, \quad [\Lambda, {\bar{\Phi}}_0] =
i{\bar{\Phi}}_0 
\label{SO3G1}
\ee
\be
[\Phi_0, {\bar{\Phi}}_0] = 2i\Lambda \,.
\label{SO3G2}
\ee
In other words, the elements $\Lambda,\Phi_0,{\bar{\Phi}}_0$ generate
an $SO(3)$ subgroup of ${\cal G}$, which we denote by 
$SO(3)_{\cal G}$. The $SO(2)_{\cal G}$ subgroup generated by $\Lambda$ is
therefore not arbitrary, but must extend to $SO(3)_{\cal G}$. 

The related reason for restricting $\Lambda$ applies in the case 
that $\Sigma = \HH^2$. Consider the action of $SO(3)$ on 
$\R^4 = \R^1 \times \R^3$. It acts in the standard way on 
the $\R^3$ factor, with 2-spheres as generic orbits. The conformal 
equivalence $\HH^2 \times S^2 \cong \R^4 - \R^1$ arises from the 
manipulation of the $\R^4$ metric,
\bea
ds^2 &=& d\tau^2 + dr^2 + r^2(d\theta^2 + \sin^2\theta \, d\varphi^2)
\label{R4metric} \\
     &\cong & \frac{2}{r^2}(d\tau^2 + dr^2) + 
2(d\theta^2 + \sin^2\theta \, d\varphi^2) \,. \label{congmetric}
\eea
The first factor in (\ref{congmetric}) is the metric on $\HH^2$ in 
the upper-half-plane model, with $r>0$, and the Gauss curvature is $-\half$. 
In terms of the complex coordinate \footnote{
We use $z=\tau+ir$ here, exchanging the role 
of $z$ and $\bar z$ compared to $z=r+i\tau$ in Ref.\cite{Witten:1976ck}. 
} $z=\tau+ir$, the metric is 
$\frac{2}{({\rm Im} z)^2}dzd{\bar z}$. Now notice that the $\tau$-axis 
of $\R^4$, where $r=0$, is excluded here. This is the excluded $\R^1$,
and it is the boundary of $\HH^2$. To have well-defined
$SO(3)$-invariant, self-dual Yang--Mills fields on all 
of $\R^4$, the $SO(3)$ invariance must hold also on this line. But
here the isotropy group jumps -- it is all of $SO(3)$. So we need to 
be able to lift $SO(3)$ to a subgroup $SO(3)_{\cal G}$ in 
${\cal G}$, and for consistency, $\Lambda$ must be
one generator of $SO(3)_{\cal G}$. In other words, in addition to
$\Lambda$, there should be two elements $\Phi_0, {\bar\Phi}_0$ of
${\mathfrak g}^*$, such that the algebra (\ref{SO3G1}) and (\ref{SO3G2}) holds.
As we saw above, this implies that the fields on $\HH^2$ can approach 
vacuum values on the boundary. The lift of these fields to 
$\HH^2 \times S^2$ can then be extended to the $\tau$-axis of $\R^4$, to 
give finite-action self-dual Yang--Mills fields on $\R^4$.

From now on, we shall suppose that $\Lambda$ is one
generator of an $SO(3)_{\cal G}$ subgroup of ${\cal G}$.

\vspace{5mm}

\noindent
{\bf 3.\ A maximally non-abelian example}

Let us now choose ${\cal G} = SU(2N)$, whose Lie algebra consists 
of $2N \times 2N$, antihermitian traceless matrices. $\Lambda$ can 
always be conjugated into the Cartan subalgebra of diagonal matrices
\be
\Lambda=i\left(
\begin{array}{cccc}
\Lambda_1& & & \\
 &\Lambda_2& & \\
 & &\ddots& \\
 & & &\Lambda_{2N}
\end{array}
\right) \,,
\ee 
with $\Lambda_{\alpha}$ real and $\sum\Lambda_{\alpha} = 0$.
To obtain a large non-abelian centralizer of $\Lambda$ and hence 
$SO(2)_{\cal G}$, we want 
as many as possible of the $\Lambda_{\alpha}$ to be equal. The constraint 
$[\Lambda, \Phi] = -i\Phi$ is satisfied by the $2N
\times 2N$ matrices $\Phi$, where the matrix element
$\Phi_{\alpha\beta}$ can be non-zero only if $\Lambda_{\beta} -
\Lambda_{\alpha} = 1$. To obtain a large non-zero part of $\Phi$, we want
as many as possible of the differences $\Lambda_{\beta} -
\Lambda_{\alpha}$ to be 1. Combining these requirements, the optimal
choice is
\be
\Lambda=\frac{i}{2}\left(
\begin{array}{cc}
{\bf 1}_N & 0 \\
0 &-{\bf 1}_N 
\end{array}
\right) \,,
\label{eq:block_matrix}
\ee  
where ${\bf 1}_N$ is the unit $N \times N$ matrix. This gives a
maximally large gauge group and Higgs field after dimensional
reduction. 

The constraints (\ref{constraints1}) and (\ref{constraints2}) are 
satisfied by fields of the form
\be
{\bf A}_z=\left(
\begin{array}{cc}
A_z & 0 \\
0 & \widetilde{A}_z 
\end{array}
\right) \,, \quad
{\bf A}_{\bar z}=\left(
\begin{array}{cc}
A_{\bar z} & 0 \\
0 & \widetilde{A}_{\bar z} 
\end{array}
\right)
\label{Ggaugepot}
\ee
\be
\Phi=\left(
\begin{array}{cc}
0 & 0\\
H & 0 
\end{array}
\right) \,, \quad
{\bar \Phi}=\left(
\begin{array}{cc}
0 & H^\dagger \\
0 & 0 
\end{array}
\right) \,,
\label{GHiggs}
\ee
where the non-zero parts are $N \times N$ blocks. The reduced gauge
group $G$ is $S(U(N) \times \widetilde{U(N)})$, i.e. $U(N) \times 
\widetilde{U(N)}$ with overall determinant 1. The Lie algebra is that 
of $SU(N) \times \widetilde{SU(N)} \times U(1)$. The notation 
$ \, \widetilde{} \,$ conveniently distinguishes the factors of the gauge
group and the corresponding gauge potentials $A$ and $\widetilde{A}$. 

There is an $SO(3)_{\cal G}$ algebra here, satisfying (\ref{SO3G1})
and (\ref{SO3G2}), with $\Lambda$ as above and
\be
\Phi_0=\left(
\begin{array}{cc}
0 & 0\\
{\bf 1}_N & 0 
\end{array}
\right) \,, \quad
{\bar \Phi}_0=\left(
\begin{array}{cc}
0 & {\bf 1}_N \\
0 & 0 
\end{array}
\right) \,.
\ee
Hence there is a zero-energy vacuum, with $H = {\bf 1}_N$, 
where the $SU(N)$ and $\widetilde{SU(N)}$ gauge groups are locked, 
instead of the colour-flavour locking mentioned in the 
introduction.

Substituting the expressions (\ref{Ggaugepot}) and (\ref{GHiggs}) into 
the generic Bogomolny equations (\ref{B1})--(\ref{B3}), we find the 
Bogomolny equations for the unconstrained fields
\bea
F_{z\bar z} &=& \frac{\sigma}{8}\left(-{\bf 1}_N + H^\dagger H \right) 
\label{Bogo1} \\
\widetilde{F}_{z\bar z} &=& \frac{\sigma}{8}\left({\bf 1}_N -
H  H^\dagger \right)
\label{Bogo2} \\
D_z H^\dagger &=& 0 \label{Bogo3} \\
D_{\bar z} H &=& 0 \,, \label{Bogo4} 
\eea
where $F, \widetilde{F}$ are the field tensors of $A,\widetilde{A}$,
respectively, and $D_z H^{\dagger} = \pr_z H^{\dagger} + A_z
H^{\dagger} - H^{\dagger}\widetilde{A}_z$, 
$D_{\bar z}H = \pr_{\bar z}H + \widetilde{A}_{\bar z}H - H A_{\bar z}$. 
These equations are gauge invariant under $G$,
with $U(N)$ acting on $H$ from the right, and $\widetilde{U(N)}$
acting from the left. So $H$ is a Higgs field in the
bifundamental representation of $G$. 

Note that if the sizes $N$ and $N'$ of the two blocks of 
matrices in eqs.~(\ref{eq:block_matrix}) and (\ref{Ggaugepot})
were unequal, the Higgs fields coming from the off-diagonal 
elements in eq.~(\ref{GHiggs}) would not be square matrices. 
By taking a trace, we can easily see that the 
corresponding Bogomolny equations (\ref{Bogo1}) and 
(\ref{Bogo2}) would then not allow the vacuum solution with 
vanishing field strengths 
${F}_{z\bar z}=\widetilde{F}_{z\bar z}=0$. 
This is another reason why we should choose the symmetric 
situation which necessitates the even size $2N$ of the starting 
unitary gauge group $SU(2N)$. 
 
\vspace{5mm}

\noindent
{\bf 4.\ Moduli matrix and master equations} 

Let us split the $U(N)$ and $\widetilde{U(N)}$ gauge potentials $A$
and $\widetilde{A}$ into their traceless $SU(N)$ and $\widetilde{SU(N)}$
parts $A^{(0)}$ and $\widetilde{A}^{(0)}$, and a common $U(1)$ part $a$.
The Bogomolny equations now take the form 
\begin{equation}
F_{z\bar z}^{(0)} + \frac{i}{2}{\bf 1}_N f_{z\bar z} =
\frac{\sigma}{8}\left(-{\bf 1}_N + H^\dagger H \right) 
\label{eq:vector_bps1}
\end{equation}
\begin{equation}
{\widetilde F}_{z\bar z}^{(0)} - \frac{i}{2}{\bf 1}_N f_{z\bar z} = 
\frac{\sigma}{8}\left({\bf 1}_N - H H^\dagger \right) 
\label{eq:vector_bps2}
\end{equation}
\begin{equation}
D_{\bar z} H = 0 \,,
\label{eq:higgs_bps}
\end{equation}
where $f_{z{\bar z}} = \pr_z a_{\bar z} - \pr_{\bar z} a_z$ 
and $D_{\bar z} H = \pr_{\bar z} H 
+ \widetilde{A}_{\bar z}^{(0)} H - H A_{\bar z}^{(0)}
- ia_{\bar z} H$. We suppress the equation (\ref{Bogo3}), as this is 
just the hermitian conjugate of (\ref{Bogo4}). By taking the traceless 
and trace parts of eqs.~(\ref{eq:vector_bps1}) and
(\ref{eq:vector_bps2}), we could decompose the Bogomolny 
equations into a set of coupled equations for the
$SU(N)$, $\widetilde{SU(N)}$ and $U(1)$ parts. For the rest of this
section we drop the superscript ${}^{(0)}$, remembering that capital $A,F$
etc. refer to $SU(N)$. 

Let us define a real gauge parameter function $\psi(z, \bar z)$ and 
$SL(N, \C)$ gauge parameter matrix functions $S(z, \bar z)$ and 
$\widetilde{S}(z, \bar z)$ by 
\begin{equation}
a_{\bar z} = -\frac{i}{2} \, \pr_{\bar z} \psi \,, 
\qquad 
A_{\bar z}=S^{-1}\partial_{\bar z} S \,, 
\qquad 
\widetilde{A}_{\bar z}=\widetilde{S}^{-1}\pr_{\bar z} \widetilde{S} \,. 
\label{eq:gauge_parameter}
\end{equation}
Using these, the Bogomolny equation (\ref{eq:higgs_bps}) for $H$ can 
be solved in terms of a holomorphic moduli matrix $H_0(z)$, as 
\cite{Eto:2005yh, Eto:2006pg, Eto:2009wq} 
\begin{equation}
H(z,{\bar z})=e^{\half\psi(z,{\bar z})} 
{\widetilde S}^{-1}(z, \bar z) H_0(z) S(z, \bar z) \,. 
\label{eq:sol_higgs_bps}
\end{equation}
By defining the gauge invariant quantities 
$\Omega \equiv S S^\dagger$ and 
$\widetilde \Omega \equiv \widetilde S \widetilde S^\dagger$, 
the matrix Bogomolny equations (\ref{eq:vector_bps1}) and 
(\ref{eq:vector_bps2}) can now be reexpressed as 
\begin{eqnarray}
\partial_z \partial_{\bar z} \psi 
 = \frac{\sigma}{4} \left( -1 + \frac{1}{N} e^{\psi} 
{\rm Tr}( \widetilde\Omega^{-1} H_0 \Omega H^\dagger_0)\right) 
 \label{eq:master_u1}
\end{eqnarray}
\begin{eqnarray}
\partial_z (\Omega^{-1} \partial_{\bar z} \Omega ) 
 = \frac{\sigma}{8} e^{\psi} \left( 
H^\dagger_0 \widetilde \Omega^{-1} H_0 \Omega
- \frac{1}{N} {\bf 1}_N {\rm Tr}( 
\widetilde\Omega^{-1} H_0 \Omega H_0^\dagger)
\right)
 \label{eq:master_un1}
\end{eqnarray}
\begin{eqnarray}
 \partial_z (\widetilde \Omega^{-1} \partial_{\bar z} \widetilde \Omega ) 
 = - \frac{\sigma}{8} e^{\psi} \left(
\widetilde \Omega^{-1} H_0 \Omega H_0^\dagger  
- \frac{1}{N} {\bf 1}_N {\rm Tr}( 
\widetilde\Omega^{-1} H_0 \Omega H_0^\dagger)
\right) \,.
 \label{eq:master_un2}
\end{eqnarray}
We call eqs.~(\ref{eq:master_u1})--(\ref{eq:master_un2}) 
the master equations for the $U(1)$, $SU(N)$ and 
$\widetilde{SU(N)}$ gauge groups, respectively. 
It has been shown 
that the solution 
of the $U(1)$ master equation (\ref{eq:master_u1}) exists 
and is unique for the given source 
${\rm Tr}( \widetilde\Omega^{-1} H_0 \Omega H^\dagger_0)$ 
\cite{Sakai:2005kz}. Similarly, we conjecture that the solution 
$\psi, \Omega, \widetilde \Omega$ of the coupled $U(1)$ and $SU(N)$ 
master equations (\ref{eq:master_u1})--(\ref{eq:master_un2}) exists 
and is unique for a given moduli matrix $H_0(z)$. 

Note that the moduli matrix is defined up to holomorphic 
gauge equivalence by $SL(N, \C)$ transformations from the left and right, 
\begin{equation}
H_0(z) \to \widetilde V(z) H_0(z) V(z) \,, \quad 
S \to V^{-1}(z) S \,, 
\quad \widetilde S \to \widetilde V(z) \widetilde S \,, 
  \label{eq:V-trans}
\end{equation}
with $V(z), \widetilde V(z)$ holomorphic in $z$, and of unit
determinant. This moduli matrix formalism is very similar to the 
case of the $U(N)$ gauge theory with $N$ flavours of 
Higgs fields in the fundamental representation 
\cite{Eto:2005yh, Eto:2006pg, Eto:2009wq}, except that here we have 
two gauge groups $SU(N), \widetilde{SU(N)}$ besides a $U(1)$ gauge 
group.  

Transposing the $SU(N)$ master equation (\ref{eq:master_un1}), 
we observe that the $\widetilde{SU(N)}$ master equation 
(\ref{eq:master_un2}) can be obtained by the transformation 
\begin{equation}
H_0 \longleftrightarrow H_0^T \,, 
\qquad 
\widetilde \Omega^{-1} \longleftrightarrow \Omega^T \,. 
\label{eq:master_eq_symmetry}
\end{equation}
The same transformation also gives (\ref{eq:master_un1}) from 
 (\ref{eq:master_un2}). 
This implies that for a symmetric moduli matrix 
$H_0=H_0^T$, the solution has the symmetry $\widetilde \Omega^{-1}
=\Omega^T$. 

On $\R^2$, where $\sigma = 1$, we cannot expect the master equations
to be integrable. However on the hyperbolic plane $\HH^2$, where $\sigma = 
\frac{2}{({\rm Im} z)^2}$, the equations are formally
integrable \cite{Leznov:1980tz,Popov}. Possibly this also applies 
to the multi-flavour $U(N)$ gauge theory on $\HH^2$, but this has 
not been established. It is interesting to observe that in the 
hyperbolic case, the explicit factor of $\sigma$ can be eliminated 
from the Bogomolny equations and the master equations 
\cite{Leznov:1980tz}. This is because
$\sigma$ satisfies the Liouville equation $\pr_z \pr_{\bar z}
(\log\sigma) = \frac{1}{4}\sigma$, and if we make the
transformation $\psi \to \psi' = \psi + \log\sigma$, the master equations
become
\begin{eqnarray}
\partial_z \partial_{\bar z} \psi' 
 = \frac{1}{4N} e^{\psi'} 
{\rm Tr}( \widetilde\Omega^{-1} H_0 \Omega H_0^\dagger) 
 \label{eq:master_u1_straighten}
\end{eqnarray}
\begin{eqnarray}
\partial_z (\Omega^{-1} \partial_{\bar z} \Omega ) 
 = \frac{1}{8} e^{\psi'} \left( 
H_0^\dagger \widetilde \Omega^{-1} H_0 \Omega
- \frac{1}{N} {\bf 1}_N {\rm Tr}( 
\widetilde\Omega^{-1} H_0 \Omega H_0^\dagger)
\right)
\label{eq:master1_straighten}
\end{eqnarray}
\begin{eqnarray}
\partial_z (\widetilde \Omega^{-1} \partial_{\bar z} \widetilde \Omega ) 
 = - \frac{1}{8} e^{\psi'} \left(
\widetilde \Omega^{-1} H_0 \Omega H_0^\dagger  
- \frac{1}{N} {\bf 1}_N {\rm Tr}( 
\widetilde\Omega^{-1} H_0 \Omega H_0^\dagger)
\right) \,.
\label{eq:master2_straighten}
\end{eqnarray}
If further, by analogy with eq.~(\ref{eq:gauge_parameter}), we define
$a'_{\bar z} = -\frac{i}{2} \, \pr_{\bar z} \psi'$, then
\begin{equation}
a'_{\bar z} =a_{\bar z} - \frac{i}{2}\partial_{\bar z}(\log\sigma) \,,
\label{eq:u1gauge_tr}
\end{equation}
and $\psi \to \psi'$, $a_{\bar z} \to a'_{\bar z}$ amounts to a 
complexified $U(1)$ gauge transformation.

\vspace{5mm}

\noindent
{\bf 5.\ Vacuum and non-abelian vortices} 

We revert here to the notation of section 3, where the
$U(N)$ gauge fields are not split up.

The vacuum of our model is given by the constant solution of 
the Bogomolny equations 
\begin{equation}
H = \left(
\begin{array}{cccc}
1 & & & \\
& 1 & & \\
& & \ddots & \\
& & & 1
\end{array}
\right)
\,, 
\quad
A=0 \,, 
\quad 
\widetilde A=0 \,.
\label{eq:vacuum}
\end{equation}
This vacuum is invariant under the diagonal gauge group 
$SU(N)_{\rm d} \,$, which is therefore the unbroken local gauge 
invariance. This contrasts with the multi-flavour $U(N)$ model, 
which is in a Higgs phase, as the gauge group is fully broken in 
the vacuum.

Exact vortex solutions are obtained using the ansatz 
\begin{eqnarray}
H=\left(
\begin{array}{cccc}
h^{(1)}& & & \\
 &1& & \\
 & &\ddots& \\
 & & &1
\end{array}
\right) \,, \quad 
A_{\bar z}= \left(
\begin{array}{cccc}
ia_{\bar z}^{(1)}& & & \\
 &0& & \\
 & &\ddots& \\
 & & &0
\end{array}
\right) \,, 
\label{eq:embedding_ansatz}
\end{eqnarray}
with $\widetilde{A}_{\bar z} = -A_{\bar z}$ so that one has an $S(U(N)
\times \widetilde{U(N)})$ gauge potential. The Bogomolny equations 
(\ref{Bogo1}) and (\ref{Bogo4}) in this case reduce to 
\begin{equation}
if^{(1)}_{z\bar z}=
\frac{\sigma}{8}\left(-1 + |h^{(1)}|^2\right)  
\label{eq:vector_bps_u1}
\end{equation}
\begin{equation}
\partial_{\bar z}h^{(1)} - 2ia^{(1)}_{\bar z}h^{(1)} =0 \,, 
\label{eq:higgs_bps_u1}
\end{equation}
where $f^{(1)}_{z\bar z} =
\partial_z a^{(1)}_{\bar z}-\partial_{\bar z}a^{(1)}_{z}$, and 
eqs.~(\ref{Bogo2}) and (\ref{Bogo3}) give nothing further.

Setting $h^{(1)} = e^{\half k + i\chi}$ with $k$ and $\chi$ real, and
eliminating $a^{(1)}_{\bar z}=(a^{(1)}_{z})^*$ 
using eq.~(\ref{eq:higgs_bps_u1}),
one finds that eq.~(\ref{eq:vector_bps_u1}) simplifies to
\be
\partial_z\partial_{\bar z} k = -\frac{\sigma}{4}(1 - e^k) \,.
\label{TaubesSam}
\ee
This is the standard gauge invariant Taubes equation for abelian 
vortices on a general surface. On the hyperbolic plane, where
$\sigma$ satisfies Liouville's equation, eq.~(\ref{TaubesSam}) itself
reduces to Liouville's equation, as first shown by Witten 
\cite{Witten:1976ck}, and its solutions have been completely worked
out in terms of Blaschke product functions. The
solutions are hyperbolic vortices and multi-vortices, that also arise 
from spherically symmetric self-dual Yang--Mills fields (i.e. instantons) in 
$SU(2)$ gauge theory on $\R^4$. 

Note that these abelian vortices embedded in 
$S(U(N)\times \widetilde{U(N)})$ gauge theory do not have full 
unit winding in the $U(1)$ subgroup of the gauge group, and they have
$SU(N)$ parts. So they are truly non-abelian. 
This situation is quite analogous to the non-abelian vortices in 
$U(N)$ gauge theories \cite{Hanany:2003hp,Auzzi:2003fs,Eto:2005yh}.  

It is clear that our construction can be extended to an arbitrary 
choice of embedding of the Witten solutions into diagonal 
elements of the $U(N)$ group, and this leads to all possible 
non-abelian vortex solutions which are restricted to lie in the
diagonal $U(1)^N$ subgroup. 

\vspace{5mm}

\vspace{.5cm}


We wish to thank David Tong for many useful discussions. 
NS thanks DAMTP for hospitality, where this work was started, 
and the Ministry of Education, Culture, Sports, Science and 
Technology, Japan for Grant-in-Aid for Scientific Research 
No. 21540279 and No. 21244036. NSM thanks the RIKEN Wako 
Institute, Japan for hospitality.


\begin{thebibliography}{99}

\bibitem{Hanany:2003hp}
  A.~Hanany and D.~Tong,
  JHEP {\bf 0307} (2003) 037
  [arXiv:hep-th/0306150].

\bibitem{Auzzi:2003fs}
  R.~Auzzi, S.~Bolognesi, J.~Evslin, K.~Konishi and A.~Yung,
  Nucl.\ Phys.\  B {\bf 673} (2003) 187
  [arXiv:hep-th/0307287].

\bibitem{Eto:2005yh}
  M.~Eto, Y.~Isozumi, M.~Nitta, K.~Ohashi and N.~Sakai,
  Phys.\ Rev.\ Lett.\  {\bf 96} (2006) 161601
  [arXiv:hep-th/0511088].

\bibitem{Witten:1976ck} E.~Witten,
  Phys.\ Rev.\ Lett.\  {\bf 38} (1977) 121.

\bibitem{Leznov:1980tz}
  A.~N.~Leznov and M.~V.~Saveliev,
  Commun.\ Math.\ Phys.\  {\bf 74} (1980) 111.

\bibitem{Popov} A.~D.~Popov,
Lett.\ Math.\ Phys.\ {\bf 84} (2008) 139
[arXiv:0801.0808 [hep-th]].

\bibitem{Taubes} C.~H.~Taubes,
Commun.\ Math.\ Phys.\ {\bf 72} (1980) 277.

\bibitem{RomST} V.~N.~Romanov, A.~S.~Schwarz and Yu.~S.~Tyupkin, 
Nucl.\ Phys.\ B {\bf 130} (1977) 209.

\bibitem{Man} N.~S. Manton,
Annals\ Phys.\ {\bf 132} (1981) 108.

\bibitem{ForMan} P.~Forg\'acs and N.~S.~Manton, 
Commun.\ Math.\ Phys.\ {\bf 72} (1980) 15.

\bibitem{Wang} H.-C.~Wang,
Nagoya\ Math.\ J.\ {\bf 13} (1958) 1.

\bibitem{LezSav} A.~N.~Leznov and M.~V.~Saveliev,
Phys.\ Lett.\  B {\bf 79} (1978) 294.

\bibitem{Bais:1978ex}
  F.~A.~Bais and H.~A.~Weldon,
  Phys.\ Rev.\  D {\bf 18} (1978) 561.

\bibitem{Wilkinson:1978zh}
  D.~Wilkinson and F.~A.~Bais,
  Phys.\ Rev.\  D {\bf 19} (1979) 2410.

\bibitem{ShiYung} M.~Shifman and A.~Yung,
Supersymmetric Solitons, Cambridge University Press, Cambridge, 2009.

\bibitem{ManRink} N.~S.~Manton and N.~A.~Rink,
Vortices on Hyperbolic Surfaces [arXiv:0912.2058 [hep-th]].

\bibitem{Gar} O.~Garc\'{\i}a-Prada,
Commun.\ Math.\ Phys.\ {\bf 156} (1993) 527.

\bibitem{Eto:2006pg}
  M.~Eto, Y.~Isozumi, M.~Nitta, K.~Ohashi and N.~Sakai,
  J.\ Phys.\ A  {\bf 39} (2006) R315
  [arXiv:hep-th/0602170].

\bibitem{Eto:2009wq}
  M.~Eto, T.~Fujimori, T.~Nagashima, M.~Nitta, K.~Ohashi and N.~Sakai,
  Phys.\ Lett.\  B {\bf 678} (2009) 254
  [arXiv:0903.1518 [hep-th]].

\bibitem{Sakai:2005kz}
  N.~Sakai and Y.~Yang,
  Commun.\ Math.\ Phys.\  {\bf 267} (2006) 783
  [arXiv:hep-th/0505136].

\end{thebibliography}
\end{document}